\documentclass[a4paper]{revtex4}
\usepackage{graphicx}
\usepackage{fancyhdr}
\usepackage{amsmath}
\usepackage{slashed}
\usepackage{subfigure}
\usepackage{epstopdf}
\usepackage{epsfig}
\usepackage{amssymb}
\usepackage{bm}
\newcommand{\beq}{\begin{equation}}
\newcommand{\eeq}{\end{equation}}
\newcommand{\ov}{\overline}
\newcommand{\pa}{\partial}

\usepackage{color}

\usepackage{color}

\begin{document}

\title{Natural SUSY from unification of SUSY breaking and GUT breaking}

%

\author{Yuji Omura}
\affiliation{Kobayashi-Maskawa Institute for the Origin of Particles and the Universe (KMI), Nagoya University, Nagoya 464-8602, Japan}

\begin{abstract}
We introduce an explicit supersymmetric unification model
where grand unified gauge symmetry breaking and supersymmetry (SUSY) breaking are caused by the same field.
Besides, the SM-charged particles are also predicted by the symmetry breaking sector, and their loop corrections induce the soft SUSY breaking terms. Especially, nonzero A-term and B-term are generated at one-loop level
according to the mediation via the vector superfields, so that the electro-weak symmetry breaking and
$125$ GeV Higgs mass could be achieved even if the stop mass is around $1$ TeV.
\end{abstract}

\maketitle

\thispagestyle{fancy}


\section{Introduction}

We are sure that there are several mysteries in our nature, although the Standard Model (SM) successfully describes 
the experimental results. For instance, the structure of the fermion generations, the hyper-charge assignment and the Higgs mass in the SM seem to suggest unknown physics behind the SM.
One of the candidates for the Beyond Standard Models (BSMs) is the supersymmetric grand unified theory (GUT),
which reveals the origin of the Higgs mass and the fermion charges.
The SM gauge group, $SU(3)_c \times SU(2)_L \times U(1)_{Y} $, is naturally unified into a $SU(5)$
gauge group, and supersymmetry (SUSY) can explain why the Higgs mass is tachyonic and around the 
electro-weak (EW) scale. The SUSY $SU(5)$-GUT has been studied for a long time, and
several issues, as well as interesting features, are pointed out so far: for instance, 
the too short life time of proton, how to realize the Yukawa coupling, and how to achieve $125$-GeV Higgs mass
as a recent topics. Futhermore, we should explicitly mention how to break SUSY and induce the soft SUSY
breaking terms in the Minimal Supersymmetric Standard Model (MSSM). 

In this paper, we propose an explicit supersymmetric GUT with $SU(5)_F \times SU(2) \times U(1)_{\phi} $ gauge
symmetry. There are two sectors in this model: the MSSM sector charged under only $SU(5)_F$, and 
the breaking sector which causes SUSY breaking and GUT symmetry breaking spontaneously.
 The SM fields belong to the MSSM sector, so that the charge quantization can be realized.
 The breaking sector consists of
one $SU(5)_F$ adjoint plus singlet filed $(\Phi)$ and 
$SU(5)_F$ fundamental and anti-fundamental fields $(\phi,~\widetilde \phi)$.
The vector-like pairs $(\phi,~\widetilde \phi)$ are also charged under
$SU(2) \times U(1)_\phi$.
In this setup, the nonzero vacuum expectation values (VEVs) of $(\Phi)$ and $(\phi,~\widetilde \phi)$ break
 $SU(5)_F \times SU(2) \times U(1)_{\phi} $ symmetry to the SM gauge groups, $SU(3)_c \times SU(2)_L \times U(1)_{Y} $, where $SU(3)_c$ is from the subgroup of $SU(5)_F$, and $SU(2)_L \times U(1)_{Y} $
are the linear combinations of the subgroup of $SU(5)_F$ and $SU(2) \times U(1)_{\phi} $.
Besides, SUSY is broken by the F-component of the part of $\Phi$. 
After the symmetry breaking, additional SM-charged particles are generated by the fluctuation
of  $\Phi$ and $(\phi,~\widetilde \phi)$ around the VEVs.
One interesting point is that  the fermionic partner of the massive gauge boson of $SU(5)_F$
could mediate the SUSY breaking effect through the gauge coupling with $\Phi$,
and play a crucial role in generating the non-zero A-term and B-term as discussed in Refs.\cite{Dermisek,Matos}.
It is well-known that large A-term could shift the upper bound on the lightest Higgs mass in 
the MSSM, even if squark is light,
 and the sizable B-term is required to realize the EW symmetry breaking.
Our A-term and B-term are given at one-loop level, so that 
they are the same order as the squark masses and gaugino masses. 
In fact, we will see that Higgs mass could be around $125$ GeV, even if the SUSY scale ($\Lambda_{SUSY}$)
is less than $O(1)$ TeV, and the B-term could be consistent with the EW symmetry breaking in our model.

The detail analysis has been done in Ref. \cite{KO}, so we briefly mention our setup 
in Sec. \ref{section2}, and discuss our results on the Higgs mass and the EW symmetry breaking in Sec. \ref{section3}. 
Sec. \ref{section4} is devoted to the summary.

\section{Setup}
\label{section2}
We briefly explain our setup of this model which causes SUSY breaking together with GUT gauge symmetry breaking,
based on Refs. \cite{KO}.

The MSSM chiral superfields are only charged under $SU(5)_F$ gauge symmetry, and they belong to
the ${\mathbf{10}}$ and $\overline{\mathbf{5}}$ representations of $SU(5)_F$, based on the Georgi-Glashow $SU(5)$ GUT \cite{GG}.
The charge assignment and matter contents are summarized in Table \ref{table1}.
${\bf 5}$-representation Higgs, $(H,\ov{H})$, are also introduced to write the Yukawa couplings in the visible sector.

\begin{table}[th]
\begin{center}
\begin{tabular}{c|cccc||ccc}
     &   ~$\ov {\bf 5}_i$~ &  ~${\bf 10}_i$~  & ~$H$~     & ~$\ov{H}$~  & ~$\phi$~ & ~$\widetilde \phi$~ &$\Phi$      \\ \hline  
$SU(5)_F$  &$\ov {\bf 5}$ &  {\bf 10} &  {\bf 5} &  $\ov{{\bf 5}}$   &  {\bf 5}    & {\bf $\ov{{\bf 5}}$}   &   {\bf adj$_5$}+{\bf 1}                  \\ 
$SU(2)$ & {\bf 1}  & {\bf 1} &{\bf 1} &{\bf 1}  & {\bf 2} &  ${\bf 2} $     &  {\bf 1}       \\ 
  $U(1)_{\phi}$  & 0  & 0 & 0 & 0   & $Q_{\phi}$ &  $-Q_{\phi}$ &  0    \\ 
\end{tabular}
\caption{
\label{table1}%
{
Chiral superfields in $SU(5)_F \times SU(2) \times U(1)$ gauge theory
}
}
\end{center}
\end{table}

$\Phi$ is the $SU(5)_F$ adjoint plus singlet field and $(\phi, \widetilde \phi)$ pair
is the vector-like under $SU(5)_F \times SU(2) \times  U(1)_{\phi}$ gauge group.
The superpotential for the breaking sector is given by 
\begin{eqnarray}
W_{SB}&=&W_R +W_{\slashed{R}}  \\
W_R&=&- h Tr_2(  \widetilde \phi \Phi \phi) + h \Lambda_G  {\rm Tr}_{5}(\Phi),  \\
W_{\slashed{R}}&=& m_{\phi} Tr_2( \widetilde \phi \phi)+c.
\end{eqnarray}
$W_R$ respects the R-symmetry and breaks the SUSY spontaneously \cite{ISS}.
$W_{\slashed R}$ breaks the R-symmetry explicitly, and gives mass of R-axion.
Some ideas to induce $W_{\slashed R}$ have been proposed in Ref. \cite{AKO1}, where the small wave-function factor
of $\Phi$ suppresses $\Phi^2$ and $\Phi^3$ terms according to the strong dynamics or the profile in the extra dimension. 
Here, we simply start the discussion from the superpotential $W_{SB}$ assuming
that such a mechanism, as discussed in Ref. \cite{AKO1}, works in underlying theories above the GUT scale, and study the symmetry breaking.  

In the global SUSY with canonical K\"ahler potential, 
the scalar potential is given by $V=|\pa_{\Phi} W_{SB}|^2+|\pa_{\phi} W_{SB}|^2+ |\pa_{\widetilde \phi}W_{SB}|^2$,
and SUSY vacua satisfy $\pa_{\Phi} W_{SB}=\pa_{\phi} W_{SB}=\pa_{\widetilde \phi} W_{SB}=0$.
In this model, $\pa_\Phi W_{SB}$ is given by
$\pa_{\Phi_{ji}} W_{SB}= -h (  \phi  \widetilde \phi)_{ij} + h \Lambda_G \delta_{ij}$ ($i,j=1, \dots, 5$),
and all elements cannot be vanishing, because $5 \times 5$ matrix $(\phi \widetilde \phi)$
has the rank $2$ \cite{ISS}.
This means that SUSY is broken by the F-components of the remnant $3$ elements in $\Phi$
and $SU(5)_F$ is also broken at the same time. There are two scales in $W_{SB}$: $m_{\phi}$ and $ \Lambda_G$.
The former relates to the $SU(5)_F$ breaking scale and the letter is concerned with the SUSY breaking scale \cite{KO}.

As well studied in Ref. \cite{KO}, the fluctuations around the SUSY and $SU(5)_F$ breaking vacuum
correspond to the fields charged under the SM gauge symmetry, whose masses are below/at the GUT scale.
Then, the extra chiral superfields works as the mediator to induce the soft SUSY breaking terms in the MSSM.
Futhermore, we could find that the massive vector superfields, which are generated by the $SU(5)_F$
symmetry breaking, could also mediate the SUSY breaking effect, because the $SU(5)_F$-adjoint field, $\Phi$,
breaks both of SUSY and GUT gauge symmetry. Then we will have some additional contributions to the soft SUSY breaking terms compared
to the ordinary gauge-mediation scenario \cite{Dermisek,Matos,KO}.
The soft SUSY breaking terms, including A-term and B-term, could be estimated as
\begin{equation}
m_{{\rm soft}} \approx \frac{ \alpha_{SM}}{4 \pi} \frac{M_p}{\Lambda_{GUT}} m_{3/2},
\end{equation}
where $M_p$, $\Lambda_{GUT}$, and $m_{3/2}$ are the Planck scale, the GUT scale and the gravitino mass respectively.
$\alpha_{SM}$ denotes the SM gauge coupling, and controls the soft SUSY breaking parameters, so that
our model has very explicit predictions for the mass spectrums of the SUSY particles.
 Blow, we summarize our prediction and especially investigate the consistency with the EW symmetry breaking
 and Higgs mass.

\section{Results}
\label{section3}
One issue in supersymmetric models is how to realize the $\mu$ and $B$ terms
which are consistent with the EW scale. Especially, the fine-tuning of $\mu$ might be required
by the recent Higgs discovery around the $125$ GeV mass region.
In fact, $125$ GeV Higgs mass seems to suggest $\Lambda_{SUSY} \gtrsim O(10)$ TeV in the simple scenarios as
discussed in Ref. \cite{Higgs1,Higgs1-2}. $O(10)$-TeV SUSY scale corresponds to at least $0.01 \%$ fine-tuning against $\mu$
without any cancellation in $m^2_{H_u}$.
As pointed out in Refs. \cite{Higgs2,Higgs3}, it is known that a special relation between $A_t$ and
squark mass relaxes the fine-tuning, maximizing the loop corrections in the Higgs mass in the MSSM.
This relation is so-called ``maximal mixing" and described as $X_t/ m_{stop} = \sqrt{6}$, where
$X_t=A_t- \mu/\tan \beta$ and $m_{stop}^2=\sqrt{m_{Q}^2 m_{U}^2}$ are defined. 
If this relation is satisfied, the $125$ GeV Higgs mass could be achieved even if the stop is less than $1$ TeV.
In our model, the predicted A-term is quite large, so that the maximal mixing might be possible. 

Unfortunately, it is difficult to realize the situation
in the minimal setup. We may have to introduce extra particles to shift the stop mass and gluino mass,
as discussed in Ref. \cite{KO}. We simply assume that the $N_{eff}$ $SU(5)_F$ vector-like pairs, which decouple at the GUT scales, contribute to the soft SUSY breaking terms, and discuss the maximal mixing, below.

\begin{figure}[!t]
\begin{center}
{\epsfig{figure=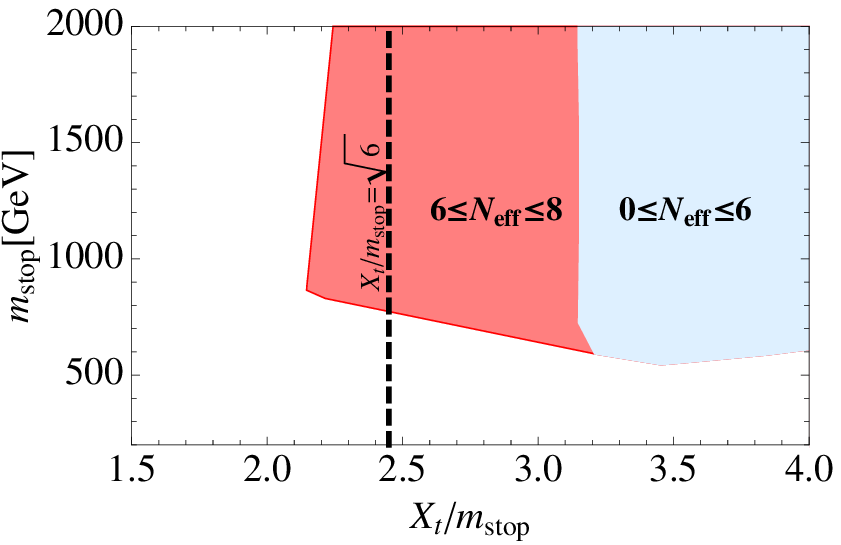,width=0.5\textwidth}}{\epsfig{figure=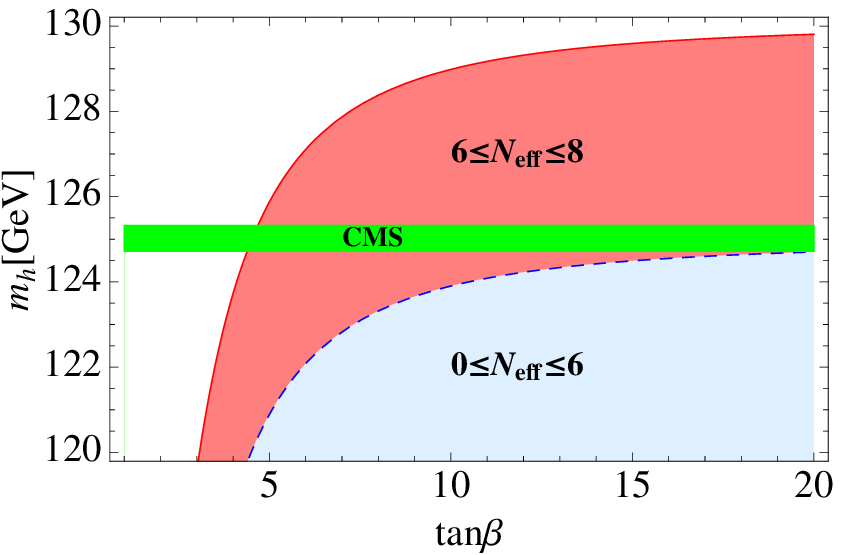,width=0.5\textwidth}}
\end{center}
\vspace{-0.5cm}
\caption{$X_t/m_{stop}$ vs. $m_{stop}$ and $\tan \beta$ vs. the lightest Higgs mass in the case with $(T_{\rm GUT},T_X)=(2 \times 10^{16} {\rm GeV},$ $10^{7} {\rm GeV})$ and $0 \leq N_{eff} \leq 6$ (light blue), $6 \leq N_{eff} \leq 8$ (light red). The dashed line corresponds to $X_t/m_{stop}=\sqrt{6}$. In the right figure, $m_h$ is calculated at the two-loop level, and $m_{stop}$ is lighter than $2$ TeV. The green band is the CMS result on Higgs mass from $h \to \gamma \gamma$, $ZZ$ channels \cite{Higgs-CMS}.
}
\label{fig3}
\end{figure}

 We can see our prediction on $X_t$ and the upper bound on the Higgs mass in the case with $0 \leq N_{eff} \leq 6$ (light blue), $6 \leq N_{eff} \leq 8$ (light red) in Fig. \ref{fig3}.
On the all regions, all masses squared of the superpartners are positive
and the GUT scale $(T_G)$ and the mass of $SU(3)_c$-adjoint field from $\Phi$ $(T_X)$ are fixed at $2 \times 10^{16} {\rm GeV}$ and $10^{7} {\rm GeV}$ respectively. 
We find that the maximal mixing could be achieved, if we allow large $N_{eff}$,
and enhances the Higgs mass, even if $m_{stop}$ is less than $1$ TeV.

\begin{figure}[!t]
\begin{center}
{\epsfig{figure=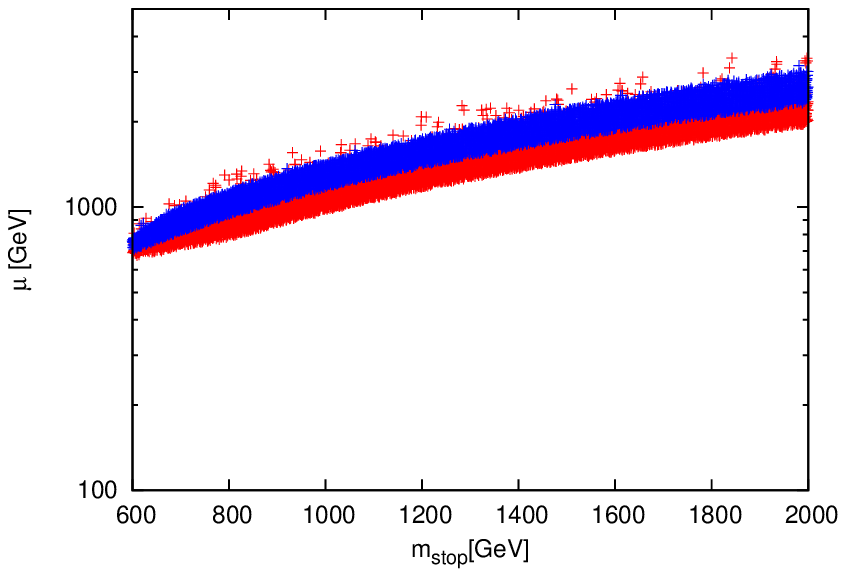,width=0.5\textwidth}}{\epsfig{figure=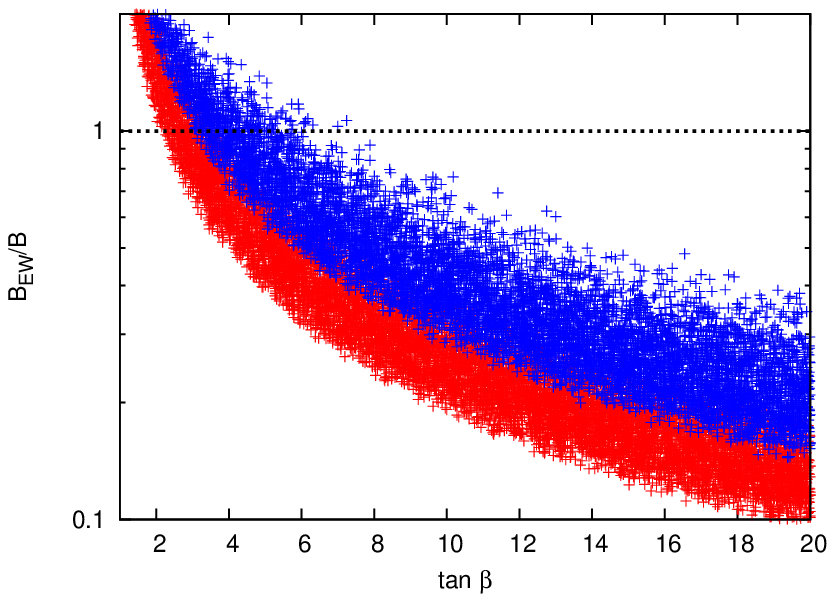,width=0.5\textwidth}}
\end{center}
\vspace{-0.5cm}
\caption{$m_{stop}$ vs. $\mu$ and $ \tan \beta$ vs. B-term in the case with $(T_{\rm GUT},T_X)=(2 \times 10^{16} {\rm GeV},$ $10^{7} {\rm GeV})$ and $0 \leq N_{eff} \leq$ (blue), $6 \leq N_{eff} \leq 8$ (red). In the right figure, $m_{stop}$ is lighter than $2$ TeV.
The dashed line is consistent with the condition for the EW symmetry breaking.
}
\label{fig4}
\end{figure}

On the other hand, we notice that there is no special cancellation in $m^2_{H_u}$ and $m^2_{H_d}$,
as we see in Fig. \ref{fig4}. Large $m_{stop}$ corresponds to large $\mu$, so that $1$-TeV squark mass
requires $1 \%$ fine-tuning against $\mu$.
The right figure in Fig. \ref{fig4} shows that small $\tan \beta$ is consistent with the EW symmetry breaking.
$B_{\rm EW}$ is the value to realize the EW symmetry breaking,
and $B$ is our prediction via the gauge mediation: $B_{\rm EW}/B$ should be unit.
It seems that $2 \lesssim \tan \beta \lesssim 6$ is necessary 
to achieve $125$ GeV Higgs mass. 
The $\tan \beta$ region may be inconsistent with the one required by $125$ GeV Higgs ($\tan \beta \gtrsim 4$)
with $m_{stop} \leq 2$ TeV.
In the appendix of \cite{KO}, we can see the parameter sets in our model, which satisfy $m_h \approx 125$ GeV and $|B_{\rm EW}/B| \approx 1$. There, $m_{stop}$ and $|\mu|$ are around $3$ TeV, and $O(0.1)$ \% fine-tuning
is required against $\mu$ term.

\section{Summary}
\label{section4}
The MSSM is one of the attractive BSMs to solve the hierarchy problem in the SM
and it may be expected to be found near future.
One big issue in the MSSM is how to control the SUSY breaking parameters,
so that many ideas and works on spontaneous SUSY breaking and 
mediation mechanisms of the SUSY breaking effects have been discussed so far.
In this paper, we proposed an explicit and simple supersymmetric model,
where the spontaneous SUSY breaking and GUT breaking are achieved
by the same sector. The origin of the hyper-charge assignment in the MSSM 
is also explained by the analogy with the Georgi-Glashow $SU(5)$ GUT \cite{GG}.
The SM-charged particles are also introduced by the
breaking sector, so that we could also predict the soft SUSY breaking terms
via the gauge mediation with the gauge and chiral messenger superfields.
The crucial role of the gauge-messenger mediation is to induce large A-terms and
B-terms at the one-loop level. We investigated the scenario with light superpartners that
such a large A-term realizes the maximal mixing and shifts the lightest Higgs mass.
In fact, we have to introduce additional contribution to the gluino mass,
but $125$ GeV Higgs mass can be achieved, even if stop is light.
$m_{stop}$ should be as light as possible to relax the fine-tuning of $\mu$ parameter.
On the other hand, the one-loop B-term can be also consistent with the EW symmetry breaking,
if $\tan \beta$ is within $2 \lesssim \tan \beta \lesssim 6$. Such small $\tan \beta$ may require large stop mass,
as we see in Figs. \ref{fig3} and \ref{fig4}.
We see that about $3$ TeV $m_{stop}$ can realize $125$ GeV Higgs mass and the EW symmetry breaking
in Ref. \cite{KO}.

Our light SUSY particles are wino, bino, and gravitino, and the mass difference is not so big.
The lightest particle is bino, and wino is heavier than bino. The mass difference is
$O(0.1) \times m_{3/2}$ GeV.
This might be one specific feature of
the gauge messenger scenario in $SU(5)$ GUT, as discussed in Ref. \cite{Bae}.

\begin{acknowledgments}
The author would like to thank T. Kobayashi for fruitful collaboration.
This work  is supported by Grant-in-Aid for Scientific research from the Ministry of Education, Science, Sports, and Culture (MEXT), Japan, No. 23104011.
\end{acknowledgments}

\bigskip 

\end{document}